\documentclass{article}

\usepackage[utf8]{inputenc}
\usepackage{natbib}
\usepackage{graphicx}
\usepackage{amsmath}
\usepackage{amssymb}
\usepackage{longtable}
\usepackage{multicol}
\usepackage{setspace}
\usepackage{mathtools}
\usepackage{rotating}
\usepackage{fontawesome}
\usepackage{qcircuit}

\usepackage{listings}
\usepackage{color}

\definecolor{dkgreen}{rgb}{0,0.6,0}
\definecolor{gray}{rgb}{0.5,0.5,0.5}
\definecolor{mauve}{rgb}{0.58,0,0.82}

\title{Quantum Honest Byzantine Agreement as a Distributed Quantum Algorithm}
\author{Marcus Edwards}
\date{July 23, 2020}

\begin{document}

\maketitle

\pagebreak

\section{Abstract}

We suggest that the Quantum Honest Byzantine Agreement (QHBA) protocol [1] essentially reduces consensus to coincidence. The volume of coincidence is the input parameter which drives a receiver to echo its input. A lack of coincidence results in no useful output from a receiver. This is a similar mechanism to the learning mechanism in cognitive modular neural architectures like Haikonen's architecture for artificial intelligence [2]. We introduce a simple feedback mechanism and quantum neuron to realize a hybrid quantum / classical machine learning network of simple nodes.

\section{Background}

The path of information between the participants in the QHBA protocol is suggestive of a fully-connected feed-forward neural network.

\vspace{5mm}

\begin{center}
    \begin{minipage}{.7\textwidth}
        \includegraphics[scale=0.25]{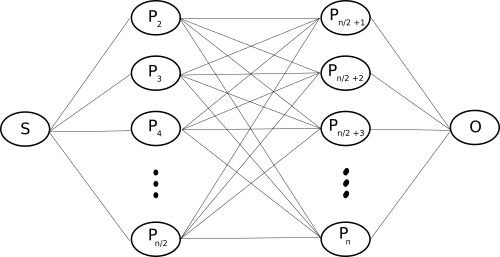}
        \vspace{5mm}
        
        The RBM-like connectivity graph of the Honest Success Byzantine Agreement
    \end{minipage}
\end{center}

\vspace{5mm}

The means of network communication between each protocol participant corresponds to a weighted I/O channel between neurons. The participants themselves correspond to neurons. The training data set is simply the sender's list $L_1$, and the output is the agreed-upon bit $B_1$.

\section{Methods}

We introduce a quantum algorithm that performs the role of the nodes in the QHBA algorithm. This algorithm defines the neurons in a learning mechanism.

\subsection{Associative Measuring Neurons}

If our network is to perform the QHBA protocol, the neurons must clearly be modified from the McCulloch-Pitts neuron. Rather than the receiver neurons simply outputting a binary classifier when the perception threshold is reached, our version of \textit{honest} receiver neurons must output its input, $B_{jk}, ID_{jk}$ when the amount of agreement or \textit{coincidence} of its inputs' values passes a threshold. This mechanism will actually suffice for the distribution neurons as well, with one small tweak. Rather than having a neuron simply output its input, we can have the neurons both \textit{measure} and then output their inputs. This will not change the function of the reveiver neurons, but will allow the distributors to achieve the replacement of 2's in the list $L_1$ provided to their inputs by the sender $S$ with probabilistically distributed 1's and 0's.

\vspace{5mm}

\noindent{\textbf{Definition 2 (Associative Measuring Neuron)}}
\textit{An associative measuring neuron will conditionally propagate a quantum state from its quantum input to its output. Its output may be a classical channel or quantum channel that will accept only basis states. Let an associative measuring neuron $k$ have the following output $y_k$ in terms of inputs $x$ from $n$ neurons, where $|X_k>$ will be a superposition of basis vectors weighted appropriately to represent their multiplicities as inputs to the neuron, and the weight of $|0^{\otimes l}>$ will represent the neuron's bias. Then the neuron achieves the non-linear effect of collapsing its output state to an that of an input which is sufficiently present so as to overcome the neuron's bias.}

\[
y_k = M |X_k> \tag{1}
\]

\vspace{5mm}

The behaviour of an associative measuring neuron is then to output the most recurring input with high likelihood, unless no input is repeated sufficiently enough, in which case the measurement will yield $|0^{\otimes l}>$ with high probability. The input coincidence threshold is manifested by $b_k$, which is the coefficient on the $|0^{\otimes l}>$ basis state and should be trainable as well as the weights $w$. A simple way to achieve this functionality would be to implement an associative measuring neuron using a series of conditionally applied quantum gates.

\vspace{5mm}

\begin{center}
    \begin{minipage}{.7\textwidth}
      \includegraphics[]{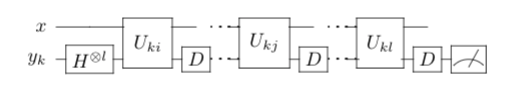}

        \vspace{5mm}
        Associative Measuring Neuron Circuit
    \end{minipage}
\end{center}

\vspace{5mm}

In this circuit, $D$ is the Grover Diffusion Operator.

\begin{center}
    \begin{minipage}{.7\textwidth}
      \includegraphics[]{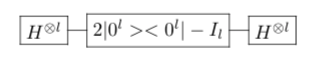}

        \vspace{5mm}
        Grover Diffusion Operator
    \end{minipage}
\end{center}

\vspace{5mm}

Each $U_{ki}$ is a circuit composed of controlled Pauli-Z phase flip gates. Each of the $U_{ki}$ is a parameterized operator that takes a set of qubit indexes $ID_{ki}$, a boolean value $B_{ki}$ and a weight $w_{ki}$. All of these parameters are classical.

If the boolean $B_{ki}$ is 1, then $U_{ki}$ is simply a multiply controlled CZ gate with all qubits in $ID_{ki}$ included as controllers in $x$ and targets in $y$. For example, if $l = 6$; $a,c \in ID_{ki}$ and $b,d,e,f,g \notin ID_{ki}$ we would have the following.

\vspace{5mm}

\begin{center}
    \begin{minipage}{.7\textwidth}
      \includegraphics[]{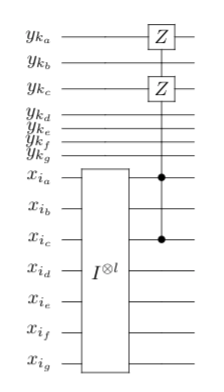}
    \end{minipage}
    
    \vspace{5mm}
    Parameterized Oracle Example with $B_{ki} = 1$
\end{center}

\vspace{5mm}

If the boolean $B_{ki}$ is 0, then a Pauli-X  gate is first applied to the input $x_i$.

\vspace{5mm}

\begin{center}
    \begin{minipage}{.7\textwidth}
        \includegraphics[]{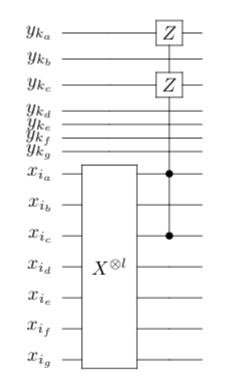}
    \end{minipage}
    
    \vspace{5mm}
    Parameterized Oracle Example with $B_{ki} = 0$
\end{center}

\vspace{5mm}

The weight $w_{ki}$ dictates the number of times the operators $U_{ki}$ and $D$ are repeated. The larger the weight with respect to other weights for other inputs to the same neuron, the more repeats. When there is only one weight and it is 1, then the operator should not be applied at all. Otherwise, $U_{ki}$ and $D$ should be repeated a number of times proportional to the number of standard deviations $N_i$ the weight $w_{ki}$ is above the mean weight $\bar{w_k}$. A negative $N$ should be reflected as well, so we will always either subtract from or add to a default number of repetitions. This default will exactly be the bias $b_k$ and will be a learned parameter itself.

\vspace{5mm}

\[
N_i = \frac{w_{ki}-\bar{w_k}}{\sqrt{\frac{1}{l} \sum_{j=1}^l (w_{kj}-\bar{w_k})^2}} \tag{2}
\]

\vspace{5mm}

The number of repetitions total will be $\left \lfloor{b_k +  N_i}\right \rfloor$.

The function of the associative measuring neuron is comparable to a number of competing Grover searches [3] performed on the same quantum state. The algorithm is designed such that the effect of a search is proportional to the multiplicity of its corresponding input list $L_k^i = x_i$, \textit{if} that list matches the paramters $ID_{ki}$ and $B_k$. In the case that $ID_{ki}$ and $B_k$ correspond correctly with $x_i$, then the effect of $U_{ki}$ is to impose a negative phase on the bits corresponding to $B_k$ in $y_k$, which effectively "tags" that state for amplitude amplification. In the case that $x_i$ does not correspond with $ID_{ki}$ and $B_k$, the controlled Z gate is not applied since not all of its controlling qubits are 1's. Hence, the bad value contributes nothing to the neuron's final output state.

$S$ should encode a 2 into $L_1$ by simply applying a Hadamard gate to the corresponding qubits in a qubit register with a size equal to the length of $L$, preparing and sending this entire qubit register $L$ to each distributor individually. With each message, $S$ sends the parameters $ID_{ki}$ and $B_k$. The effects of the associative measuring neuron's operations will be to exactly replace any 2's in the list $L_1$ provided to their inputs by the sender $S$ with probabilistically distributed 1's and 0's via measurement of the corresponding single qubit states which will be $\frac{|0> + |1>}{\sqrt{2}}$ or $\frac{|0> - |1>}{\sqrt{2}}$ depending on whether $B_k$ is 1 or 0. The algorithm will leave the rest of the state unchanged and simply measure it.

The receivers will also be associative measuring neurons and perform the same process. However, it will be assumed that they are more likely to have conflicting inputs, and that their weights will not all agree. Also, receivers will receive their inputs $x$ from distributors and other receivers, but the parameters $ID_{ki}$ and $B_k$ will still be provided directly by $S$. The receivers should have a final state that approximates the following.

\vspace{5mm}

\[
|X_k> = b_k |0^{\otimes l}> + \sum_{i} \sum_{j \neq i} (<x_i|x_j> w_{ki}w_{kj}) |x_i> \tag{3}
\]

\vspace{5mm}

The weights and biases will be naturally normalized.

\vspace{5mm}

\[
b_k^2 + \sum_{i} \sum_{j \neq i} \delta(x_i, x_j) (w_{ki}w_{kj})^2 = 1 \tag{4}
\]

\vspace{5mm}

While the machine begins with a nearly equal number of distributors and receivers, the neurons which do not receive consistent data do not output information and their input sources are considered unreliable. This decreases the number of useful neurons in the network and may reduce the size of either of the two layers. This is how the QHBA selects trusted paths of information through the network. This is not dissimilar to the way that pathways between neurons are created through learning in Haikonen's cognitive modular neural network.

\subsection{Training Via Quantum Binding Commitment}

We can make use of an optimally secure mechanism called a Quantum Binding Commitment [5] to define the training data for the system. In our network, we will want the consensus to result in agreement on a single bit. So, the measure of fidelity used in our training is trivial: a single bit (a sender's vote) that is known initially only by the sender and the training code which can be a very simple, visible, immutable and infinitely running script.

\vspace{5mm}
\begin{center}
\includegraphics[scale=0.2]{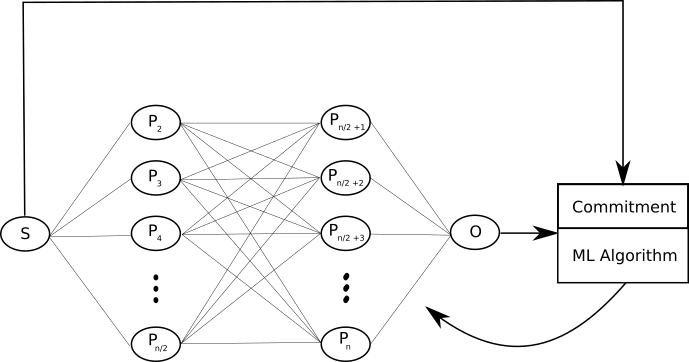}

Quantum Bit Commitment Training
\end{center}
\vspace{5mm}

\section{Calculations}

It is an intended feature of the associative measuring neuron's design that if the size of input lists and outputs is $l = 6$, its behaviour can be realized today using IBM's commercially available Q System One or IBM Q 16 Melbourne system which is free to use for research purposes. Some restrictions must be applied to the neuron in order to ensure that it can be implemented using either system, since the superconducting Transmon qubit networks of these systems are not fully-connected. In order to make the associative measuring neuron compatible with the Melbourne, a sender simply must choose $ID_{ki}$ and $B_k$ that specify a controlled Z operation that is possible to implement. Many control configurations can be achieved using qubit swapping.

\vspace{5mm}
\begin{center}
\includegraphics[scale=0.7]{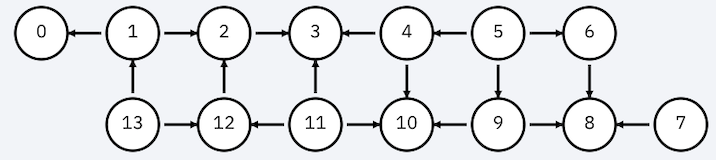}

IBM Q 16 Melbourne Connectivity Graph
\end{center}
\vspace{5mm}

The number of $D$ and $U_{ki}$ operations applied, the less fidelity we will have in the neuron's outputs due to decoherence. A minimum fidelity should be chosen and used to select the range of possible values that will be taken by the default repetition bias $b_k$. This fidelity can be dynamically chosen based on the calibration parameters of the Melbourne, for example, which fluctuate but are available at a given time. The average $T1$ and $T2$ times for the IBM Q System One are reportedly $74 \mu s$ and $69 \mu s$ respectively [6]. The Melbourne's decoherence times are similar but vary depending on the qubits involved, as evidenced by the figure. IBM reports that the average decoherence times for the Melbourne are $T1 = 67.50 \mu s$ and $T2 = 22.40 \mu s$ [7]. We only will consider the Melbourne's limitations thoroughly, since it is less advanced and more limited than the IBM Q System One.

\vspace{5mm}
\begin{center}
\includegraphics[scale=0.4]{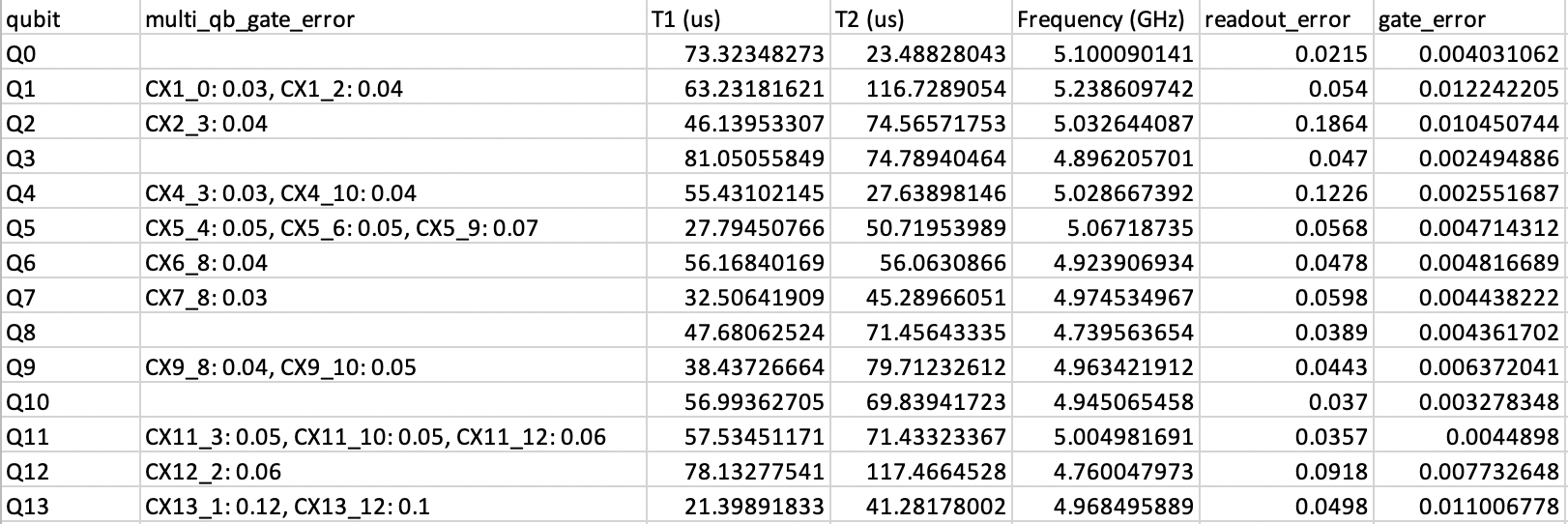}

IBM Q 16 Melbourne Calibration Details July 17th 2019
\end{center}
\vspace{5mm}

The gate times of the Melbourne are updated continuously and published publicly [8], and the average amount of time required for a CX gate is around 350ns.

\vspace{5mm}
\begin{center}
\includegraphics[scale=0.4]{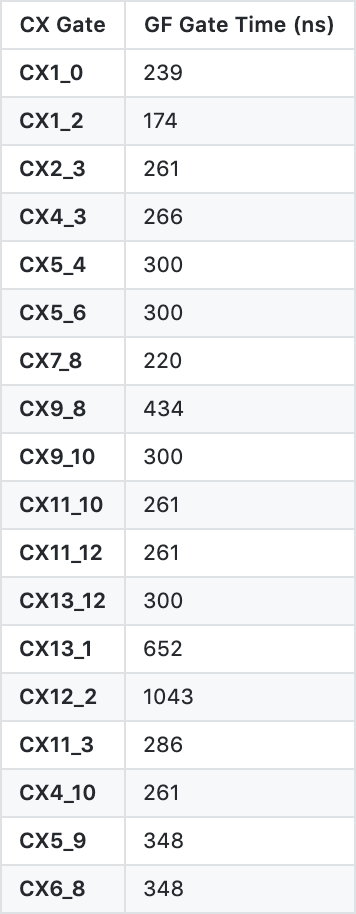}

IBM Q 16 Melbourne Gate Time Details August 7th 2019
\end{center}
\vspace{5mm}

A CZ gate is realized in the IBM system using a CX and two single qubit Hadamard gates. Each $U_{ki}$ is a pair of controlled CCZ gates, and either 0 or 6 X gates. A CCZ gate can be realized using CNOT, $T^\dagger$, and $T$ gates via an optimal decomposition [10]. This requires six CX gates. The Grover diffusion operator can be realized using Hadamard gates surrounding a multiply controlled Z operation as well, as per [9].

\begin{center}
    \begin{minipage}{.7\textwidth}
      \includegraphics[]{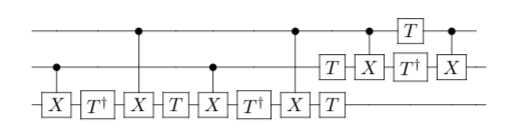}
        \vspace{5mm}
        CCZ Gate Acheivable Using IBM Q
    \end{minipage}
\end{center}

\vspace{5mm}

Generally, the Grover diffusion operator would involve a multiply controlled Z gate with a number of controls equal to the size of the output register, minus one. We can get away with simplifying the Grover diffusion operator for our case by realizing that the operator will only ever be used to rotate the state towards basis states with two non-zero qubit values. During each such rotation, the diffusion operator can be realized by a single CZ gate which involves the two qubits that correspond to the particular input $x_i$'s corresponding indexes $ID$. This will rotate the high dimensional output state towards the target basis state in the relevant degrees of freedom, and leave the other degrees of freedom untouched. However, each input $x_i$ may adjust the overall output state in different degrees of freedom, and together rotate the state in any arbitrary direction. Using this approach, the Grover diffusion operator can be realized using six single qubit gates and one CX gate.

The single qubit gates involved in the algorithm each have a time penalty as well. These time penalties can be understood by decomposing each unitary gates into its set of actual physical gates that are used to implement them in IBM's system. IBM's computers support three types of single qubit gates, the first two (u1, u2) are relevant for us:

\[
  u1(\lambda) = \begin{bmatrix}
        1 & 0 \\
        0 & e^{\lambda i}
    \end{bmatrix}
\]
\[
  u2(\phi, \lambda) = \frac{1}{\sqrt{2}} \begin{bmatrix}
        1 & -e^{\lambda i} \\
        e^{\phi i} & e^{(\phi i + \lambda i)}
    \end{bmatrix}
\]

\vspace{5mm}

Any single qubit gate which has the form given by $u1$ is implemented using Frame Change (FC) operation, which does not physically take any time but actually influences the frame of the following operation and takes no time in and of itself. We can see that the $T$ and $T^{\dagger}$ gates do have the corresponding form.

\[
T = \begin{bmatrix}
        1 & 0 \\
        0 & e^{\frac{\pi}{4} i}
    \end{bmatrix}
= u1(\frac{\pi}{4})
\]

\[
T^{\dagger} = \begin{bmatrix}
        1 & 0 \\
        0 & e^{-\frac{\pi}{4} i}
    \end{bmatrix}
= u1(-\frac{\pi}{4})
\]

The Hadamard gate is also used in our algorithm, and matches the form of u2. Any gate which has the form of u2 is implemented using a physical Gaussian-Derivative (GD) pulse parameterized by two frame changes. A GD pulse takes 60ns itself, and invokes a 10ns buffer time.

\[
H = \frac{1}{\sqrt{2}} \begin{bmatrix}
        1 & 1 \\
        1 & -1
    \end{bmatrix}
= u2(2\pi, 3\pi)
\]

The final type of gate that is relevant for our work is the CX gate, which makes use of both FC and GD physical gates as well as Gaussian Flattop (GF) pulses. We have already addressed the time requirements for CX gates depending on the qubits involved.

Understanding this, we may say that the time requirement for $U_{ki}$ is at most equivalent to that of two CCZ gates and six X gates.

\vspace{5mm}

\[
T(U_{ki}) \; \dot{=} \; 2 \cdot 0ns + 6 \cdot 350ns = 2100ns \tag{5}
\]

\vspace{5mm}

Similarly, the time requirement for our simplified Grover diffusion operator is that of four Hadamard gates, two X gates and one CX gate.

\vspace{5mm}

\[
T(D) \; \dot{=} \; 2 \cdot 0ns + 4 \cdot 70ns + 1 \cdot 350ns = 640ns \tag{6}
\]

\vspace{5mm}

The overall time cost of a repetition of $DU_{ki}$ is then given by equation (7).

\vspace{5mm}

\[
T_{rep} = T(U_{ki}) + T(D) = 2740ns \tag{7}
\]

\vspace{5mm}

The time requirement for an associative measuring neuron's operation in its entirety will then be given by equation (8).

\vspace{5mm}

\[
T_{assoc} = \sum_i \left \lfloor{b_k + N_i}\right \rfloor \cdot 2740ns \tag{8}
\]

\vspace{5mm}

To ensure this operation completes within an acceptable window, we simply enforce that $T_{assoc} < T2$. The most expensive associative measuring neuron operation will involve $\frac{|P|}{2}$ inputs $x_i$. So, for example a system with ten participants would yield a maximal $T_{assoc}$ time of $max(T_{assoc} | \; |P|)$.

\vspace{5mm}
\[
max(T_{assoc} | \; |P|) = \sum_{i=0}^{\frac{|P|}{2}} \left \lfloor{b_k + N_i}\right \rfloor \cdot 2740ns 
\]
\vspace{5mm}

In this scenario, $N_i$ would be standard deviations of each $\frac{|P|}{2}$ points. To keep an associative measuring neuron operation under the shortest time constraint, which is $T2 = 22.40 \mu s$ on the Melbourne, we must limit either the number of participants in the network $|P|$, or cap the number of standard deviations $N_i$ at some maximum range. It is more appealing for the machine learning algorithm to take the number of participants as a parameter and adjust the range of the maximum considered standard deviation. So, we can define a maximum range $max(N_i |\; |P|)$.

\vspace{5mm}

\[
T2 = max(T_{assoc} | \; |P|) = \sum_{i=0}^{\frac{|P|}{2}} \left \lfloor{b_k + max(N_i)}\right \rfloor \cdot 2740ns
\]

\[
22.40 \mu s = \sum_{i=0}^{\frac{|P|}{2}} \left \lfloor{b_k + max(N_i)}\right \rfloor \cdot 2740ns 
\]

\[
\frac{22.40 \mu s}{2740 ns} = \sum_{i=0}^{\frac{|P|}{2}} \left \lfloor{b_k + max(N_i)}\right \rfloor
\]

\[
\frac{22.40 \mu s}{2740 ns} - \sum_{i=0}^{\frac{|P|}{2}} b_k \; \dot{=} \; \sum_{i=0}^{\frac{|P|}{2}} max(N_i)
\]

\[
\frac{22.40 \mu s}{2740 ns} - \sum_{i=0}^{\frac{|P|}{2}} b_k \; \dot{=} \; {\frac{|P|}{2}} max(N_i)
\]

\[
max(N_i |\; |P|) \; \dot{=} \; \frac{2}{|P|} \cdot \frac{22.40 \mu s}{2740 ns} - \sum_{i=0}^{\frac{|P|}{2}} b_k
\]

\vspace{5mm}

In the worst case, $\sum_{i=0}^{\frac{|P|}{2}} b_k \rightarrow \frac{|P|}{2}$ since $0 \leq b_k \leq 1$.

\vspace{5mm}

\[
max(N_i |\; |P|) \; \dot{=} \; \frac{2}{|P|} \cdot \frac{22.40 \mu s}{2740 ns} - \frac{|P|}{2}
\]

\vspace{5mm}

The system will become functionally useless when $max(N_i |\; |P|)$ approaches 0. Therefore we can conclude that the system will be able to handle only 6 participants if our quantum associative measuring neurons were used at each node today.

Also, if this system were implemented today, each participant would not have a local quantum computer to use for their associative measuring neuron operations. Rather, they would need to delegate their quantum computations to a central quantum computer. Today, the best option would be IBM's system. The amount of time spent in the queue waiting for each others' operations to complete would render the speedup from using Grover's search pointless.

\section{Discussion}

Despite the conclusion that this system is not practical to implement today, the work we did in the last subsection gives us a method for predicting how useful the system will be in the future, when we have access to better quantum computers.

IBM claims that they intend to eventually improve their coherence (T2) times to 1-5 milliseconds, and suggest that they are exponentially approaching this goal according to a relationship similar to Moore's law for integrated electronics [6]. Assuming this goal is reached within the next decade, which is generally considered to be feasible with at least a non-zero possibility, our scheme would be able to support roughly 85 participants in each vote.

The true randomness of the probabilistic outcomes from measuring the states $\frac{|0> + |1>}{\sqrt{2}}$ and $\frac{|0> - |1>}{\sqrt{2}}$ is a valuable cryptographic asset when a quantum associative measuring neuron is used for a security protocol due to the outcome being truly random. Also, the Grover's search algorithm provides a known quadratic speedup over equivalent classical methods, when the number of applied operations is compared to the number of classical records checked for the value searched for [4].

However, it is important to point out that an associative measuring neuron with a limited repetition capacity can be easily classically simulated. So, the entire system described thus far could theoretically be replaced with a classical equivalent. This would mean that we do not gain the security and efficiency benefits of the quantum algorithms employed. However, it would mean that scaling the system to support any arbitrarily large number of users would be possible.

An optimal network scheme would incorporate both quantum and classical elements to take advantage of as much quantum security and speedup as possible with the resources available whilst also supporting an arbitrary number of users. 

The machine learning inspired element of the consensus algorithm implemented at any scale would be beyond the capabilities of any quantum computing technology that exists today. However, it would be well within the reach of modern classical technology. So, we will assume that it is purely classical for the forseeable future. However, it would be interesting for a future work to look into how quantum machine learning might increase the efficiency of this component as well.

On the other hand, quantum-secure communication channels are already being established and demonstrated. We posit that quantum networks will also be available for practical use in the near-term, and we can expect to use these as a resource. The ability to perform a Hadamard gate, transmit and measure the resulting state is already quite feasible. So, we can assume that at least some of the channels used by the sender of any vote can benefit from the pure randomness of the quantum approach for encoding 2's into the lists $L_{1k}$ [11].

Our system does not make assumptions on the number of participants who will be interested in participating in any given vote. Therefore it is conceivable that votes involving small numbers of people ($<6$ today, $<85$ eventually) could occur, and benefit fully from the quadratic speedup of the Grover's search. Larger votes could also occur, which would involve strictly classical neuron operations and simply trade efficiency for scalability.

The ability to use a mixture of quantum and classical channels and neurons is also an advantage. It enables this blockchain scheme to be viable throughout transitions in networking and computing technology.

\section{References}

\begin{itemize}

\item[1.] Sun, Xin, et al. “A Simple Voting Protocol on Quantum Blockchain.” International Journal of Theoretical Physics, vol. 58, no. 1, 2019, pp. 275–281.
    
\item[2.]  An Artificial Cognitive Neural
System Based on a Novel Neuron Structure and a Reentrant Modular Architecture with Implications to Machine Consciousness

\item[3.] Grover, Lov K. “A Fast Quantum Mechanical Algorithm for Database Search.” Proceedings of the Twenty-Eighth Annual ACM Symposium on Theory of Computing - STOC 96, 1996, doi:10.1145/237814.237866.

\item[4.] Bennett, C. H., Bernstein, E., Brassard, G., \& Vazirani, U. (1997). Strengths and Weaknesses of Quantum Computing. SIAM Journal on Computing, 26(5), 1510-1523. doi:10.1137/s0097539796300933

\item[5.] Unruh, Dominique. “Computationally Binding Quantum Commitments.” Advances in Cryptology – EUROCRYPT 2016 Lecture Notes in Computer Science, 2016, pp. 497–527., doi:10.1007/978-3-662-49896-5\_18.

\item[6.] “Cramming More Power Into a Quantum Device.” IBM Research Blog, 15 Mar. 2019, www.ibm.com/blogs/research/2019/03/power-quantum-device/.

\item[7.] “Quantum Devices \& Simulators.” IBM Q, 5 June 2018, www.research.ibm.com/ibm-q/technology/devices/\#ibmq\_16\_melbourne.

\item[8.] Qiskit. “Qiskit/Ibmq-Device-Information.” GitHub, github.com/Qiskit/ibmq-device-information/blob/master/backends/melbourne/V1/version\_log.md\#gate-specification.

\item[9.] Coles, et al. “Quantum Algorithm Implementations for Beginners.” 2018.

\item[10.] Shende, Vivek V., and Igor L. Markov. “On the CNOT-Cost of TOFFOLI Gates.” 2008, pp. Quant.Inf.Comp. 9(5–6):461–486 (2009).

\item[11.] Hussein Abulkasim, Atefeh Mashatan, Shohini Ghose. "Quantum-based privacy-preserving sealed-bid auction on the blockchain." Optik, Volume 242, 2021, 167039, ISSN 0030-4026, https://doi.org/10.1016/j.ijleo.2021.167039.

\end{itemize}

\end{document}